\documentclass[aps,amsfonts,amsmath,prd,preprint,nofootinbib]{revtex4}
\pdfoutput=1
\newcommand{\beq}{\begin{equation}}
\newcommand{\eeq}{\end{equation}}
\usepackage{graphicx}
\usepackage{natbib}

\usepackage{xcolor}

\usepackage{tabularx}
\usepackage{multirow}
\usepackage{booktabs}

\usepackage{amsmath}

\DeclareMathOperator\erf{erf}
\DeclareMathOperator\arcsinh{arcsinh}

\newcommand{\be}{\begin{equation}}
\newcommand{\ee}{\end{equation}}
\newcommand{\bea}{\begin{eqnarray}}
\newcommand{\eea}{\end{eqnarray}}
\newcommand{\bi}{\begin{itemize}}
\newcommand{\ei}{\end{itemize}}

\begin{document}

\title{Probability of the Initial Conditions \\ 
for Inflation and Slow Contraction}
\author{Mark P. Hertzberg\footnote{mark.hertzberg@tufts.edu}, Daniel Jim\'enez-Aguilar\footnote{Daniel.Jimenez\_Aguilar@tufts.edu}}
\affiliation{Institute of Cosmology, Department of Physics and Astronomy, Tufts University, Medford, MA 02155, USA}

\begin{abstract}
Some recent studies based on numerical relativity simulations claim that slow contraction/ekpyrosis is strongly preferred over inflation as the smoothing mechanism that brought the universe into the homogeneous, isotropic and flat state we observe today on large scales. In this paper, we evaluate the likelihood of the initial conditions employed in the aforementioned simulations by estimating the probability that a free scalar field dominating the universe at the beginning of inflation or ekpyrosis will be sufficiently homogeneous on scales comparable to the Hubble radius at that time. We explore the space of parameters that characterize the initial power spectrum of the scalar field, finding that either can be more likely than the other for a fixed choice of parameters. On the other hand, when we extremize over these parameters, we find that the maximal probability for inflation is much higher than that of ekpyrosis.
\end{abstract}

\maketitle


\tableofcontents

\newpage

\section{Introduction}

One of the most important challenges for any cosmological model is to explain the extraordinary homogeneity, isotropy and flatness of the universe on large scales. It is widely accepted today that such a smooth state can be achieved dynamically through an early stage of quasi-exponential expansion called inflation \cite{Guth:1980zm, Starobinsky:1979ty, 1980ApJ...241L..59K, 10.1093/mnras/195.3.467, Linde:1981mu, PhysRevLett.48.1220}. In the most elementary descriptions of this process, the accelerated expansion is driven by a scalar field (the inflaton) that slowly rolls down its potential. The dynamics of the quantum fluctuations of the inflaton in this framework leads to an economic and consistent explanation for the cosmic microwave background anisotropies and large-scale structure formation \cite{SDSS:2003eyi,Guth:2005zr, refId0}.  

Despite its simplicity and success, some of the features of inflation, especially those concerning the initial conditions, have been a topic of controversy over the years \cite{Penrose:1988mg,Steinhardt:2011zza,Ijjas:2013vea,Guth:2013sya,Guth:2013epa,Brandenberger:2012aj,Brandenberger:2016uzh}. One of the issues is that inflation is geodesically incomplete into the past \cite{Borde:2001nh}, so it does not provide a resolution to the initial singularity problem. Another one comes from its eternal nature \cite{Steinhardt:1982kg,Vilenkin:1983xq,Guth:2007ng}, often viewed as problematic because the set of spacetime regions where inflation ends is effectively a set of measure zero, and thus smoothness and flatness are not the general outcome. The major concern is, perhaps, whether inflation can smooth out the highly curved and warped spacetime expected to emerge from a Big Bang. According to recent work \cite{Cook:2020oaj, Ijjas:2023dnb, Garfinkle:2023vzf,Ijjas:2024oqn} supported by numerical relativity simulations, inflation will tend to end too soon or will not even start in the presence of significant inhomogeneities and anisotropies in the initial state. However, there is currently no consensus on these claims. Indeed, other groups have found that (at least large-field) inflation is generically robust to large initial spatial inhomogeneities. Some numerical studies showing this are \cite{Kurki-Suonio:1993lzy,East:2015ggf,Clough:2016ymm,Clough:2017efm,Bloomfield:2019rbs,Aurrekoetxea:2019fhr,Joana:2020rxm,Joana:2022uwc,Corman:2022alv,Elley:2024alx,Joana:2024ltg} (see also sections 3.2 and 3.3 of the review \cite{Aurrekoetxea:2024ypv}), while \cite{Kleban:2016sqm,Creminelli:2019pdh,Creminelli:2020zvc,Wang:2021hzv}, for instance, provide a more analytic perspective that leads to the same conclusions (see also \cite{Azhar:2022yip}).

The studies \cite{Cook:2020oaj, Ijjas:2023dnb, Garfinkle:2023vzf,Ijjas:2024oqn} show that a slow contraction\footnote{In this paper, we will use the terms ``slow contraction'' and ``ekpyrosis'' interchangeably.} epoch successfully brings the universe into a spatially flat Friedmann-Lemaître-Robertson-Walker (FLRW) state for generic initial conditions. This contraction phase requires a scalar field with equation of state parameter $w=p/\rho\gg1$ evolving under the influence of a negative potential. In this case, the Hubble radius shrinks much faster than the scale factor, thus solving the flatness problem in the opposite way as inflation does. Eventually, slow contraction ends and transitions to an expanding phase through a cosmic bounce \cite{Ijjas:2018qbo, Khoury:2001wf, Buchbinder:2007tw, Levy:2015awa} that preserves the smoothness achieved during the contraction. This picture succeeds in resolving the initial singularity problem, in generating a nearly scale-invariant spectrum of curvature perturbations \cite{Ijjas:2020cyh,Ijjas:2021ewd}, and also in smoothing the universe on scales even larger than the ones set by causality \cite{Ijjas:2021gkf}. However, it requires some modification of Einstein gravity or some form of energy that violates the null energy condition at the bounce \cite{Ijjas:2016vtq,Tukhashvili:2023itb}.

In \cite{Ijjas:2024oqn}, the authors compare the smoothing capabilities of inflation and ekpyrosis for several sets of initial conditions for the scalar field and the spacetime geometry. They conclude that inflation can smooth the universe to the desired degree only when the initial conditions are not too far from a FLRW universe, whereas ekpyrosis achieves successful smoothing for a much wider range of initial conditions, including nonperturbative deviations from FLRW geometry. The authors clearly state that the inflation simulations are different from the slow contraction ones in many respects, including the scalar potential and the time slicing involved in the numerical formulation of the problem. Consequently, the characteristic length scales in the initial state for each type of simulation are vastly different: the analogue of the Hubble radius in the initial-time hypersurface is of the order of the Hubble radius today for the slow contraction simulations, while it is more than 60 orders of magnitude smaller for the inflation simulations. Accordingly, the typical wavelength of the scalar field modes in each type of simulation has roughly the same relative factor. 

In this work, our goal is to estimate the likelihood of these very different initial configurations. However, we will only focus on the energy content of the initial state. In principle, it might be the case that ekpyrosis is the preferred smoothing mechanism at the cost of requiring much more unlikely initial conditions than inflation. In other words, it is possible that initial homogeneity of the scalar field on scales comparable to the Hubble radius today (as required to some extent for successful ekpyrosis) is much less likely than homogeneity on much shorter scales of the order of the Hubble radius at the beginning of inflation (as required for successful inflation), even if the homogeneity restriction is less stringent in the former case. Therefore, the specific question we want to address is the following: what is the probability that the scalar field is sufficiently homogeneous on scales comparable to the horizon at the beginning of inflation and slow contraction? 

The paper is structured as follows. In section \ref{sec:initial conditions}, we describe the initial conditions that we will be considering for the onset of inflation or ekpyrosis. In section \ref{sec:probability estimates}, we compute the aforementioned probability under certain assumptions and simplifications which will be detailed in the main text. Finally, we present the results for two different types of initial scalar field spectra in sections \ref{sec:results thermal spectrum} and \ref{sec:results cutoff spectrum}, as well as our conclusions in section \ref{sec:conclusions}. We also provide a simplified characterization of the geometry of the initial-time hypersurface in Appendix \ref{sec:appendix}.

We will use the $(-++\,\,+)$ convention for the spacetime metric and natural units ($c=\hbar=1$). Newton's gravitational constant is thus given by the square of the Planck length, or the inverse of the Planck mass squared: $G=L_{p}^2=M_{p}^{-2}$. 


\section{Initial conditions}
\label{sec:initial conditions}

We will assume that the universe is dominated by a free scalar field $\varphi$ minimally coupled to gravity at the beginning of inflation or slow contraction. The action is 
\begin{equation}
S=\int d^{4}x\sqrt{-g}\left[\frac{R}{16\pi G}-\frac{1}{2}g^{\mu\nu}\partial_{\mu}\varphi\partial_{\nu}\varphi-V(\varphi)\right]\,,
\label{eq:action}
\end{equation}
where $R$ is the Ricci scalar and $V(\varphi)$ is the scalar field potential. 

In general we are interested in inhomogeneous field fluctuations and therefore an associated inhomogeneous metric. However, for the purpose of the equations of motion, let us consider all this in reference to some ensemble-averaged metric. The specific ensembles that we will consider will be described below. Let $\delta\varphi$ be the fluctuation of the field away from the minimum of its potential. After redefining the fluctuations as $\xi=a\,\delta\varphi$ (where $a(t)$ is the ensemble-averaged scale factor) and using conformal time (defined through $d\eta=dt/a$) as the time variable, the equation of motion to lowest order in $\xi$ reads
\begin{equation}
\xi''-\nabla^{2}\xi+M_{\text{eff}}^{2}(\eta)a^{2}(\eta)\,\xi=0\,,
\label{eq:eom}
\end{equation}
where primes denote differentiation with respect to the conformal time $\eta$ and the Laplacian $\nabla^{2}$ involves spatial derivatives with respect to comoving coordinates. 
Here the effective mass $M_{\text{eff}}(\eta)$ is given by
\begin{equation}
M_{\text{eff}}^{2 }(\eta)=M^{2}-\frac{\langle R(\eta)\rangle}{6}+\frac{\langle k\rangle}{a^{2}(\eta)}\,,
\label{eq:effective mass}
\end{equation}
where $M$ is the ``bare mass'' of the field, $\langle R(\eta)\rangle$ is the ensemble average of the Ricci scalar and $k$ is the FLRW curvature index (not to be confused with the momentum $\vec{k}$ appearing below). 
Using a kind of ensemble average here, allows for the system to be {\em linear} in $\xi$. Ideally, one should insert the full self-consistent value for $R$ here, which depends on $\xi$, but that renders the full system non-linear and makes the analysis much more complicated. So we restrict to the average here as a first pass at this tough problem.

With this simplification, the solution to the equation of motion (\ref{eq:eom}), evaluated here at the initial time $\eta=\eta_{i}$, admits a mode expansion of the form
\begin{equation}
\xi(\eta_{i},\vec{r})=\int \frac{d^{3}k}{\sqrt{2\left(2\pi\right)^{3}\omega_{\vec{k}}}}\left(\bar\alpha_{\vec{k}}e^{i\vec{k}\cdot\vec{r}}+\bar\alpha^{*}_{\vec{k}}e^{-i\vec{k}\cdot\vec{r}}\right)=\int d^{3}k\,\tilde{\xi}(\vec{k})e^{i\vec{k}\cdot\vec{r}}\,,
\label{eq:xi expansion}
\end{equation}
\begin{equation}
\xi'(\eta_{i},\vec{r})\equiv\pi(\eta_{i},\vec{r})=-i\,a_{i}\int d^{3}k\sqrt{\frac{\omega_{\vec{k}}}{2\left(2\pi\right)^{3}}}\left(\bar\alpha_{\vec{k}}e^{i\vec{k}\cdot\vec{r}}-\bar\alpha^{*}_{\vec{k}}e^{-i\vec{k}\cdot\vec{r}}\right)=\int d^{3}k\,\tilde{\pi}(\vec{k})e^{i\vec{k}\cdot\vec{r}}\,,
\label{eq:pi expansion}
\end{equation}
where $a_{i}=a(\eta_{i})$ and
\begin{equation}
\omega_{\vec{k}}=\sqrt{M_{\rm eff}^{2}(\eta_{i})+\vec{k}^{2}}\,.
\label{eq:omega}
\end{equation}
Here, $|\vec{k}|=2\pi/\lambda$, with $\lambda$ the physical wavelength of the corresponding mode. On the other hand, the Fourier transforms $\tilde{\xi}(\vec{k})$ and $\tilde{\pi}(\vec{k})$ can be easily shown to be
\begin{equation}
\tilde{\xi}(\vec{k})=\frac{1}{\sqrt{2\left(2\pi\right)^{3}\omega_{\vec{k}}}}\left(\bar\alpha_{\vec{k}}+\bar\alpha^{*}_{-\vec{k}}\right)\,,
\label{eq:f xi}
\end{equation}
\begin{equation}
\tilde{\pi}(\vec{k})=-i\,a_{i}\sqrt{\frac{\omega_{\vec{k}}}{2\left(2\pi\right)^{3}}}\left(\bar\alpha_{\vec{k}}-\bar\alpha^{*}_{-\vec{k}}\right)\,,
\label{eq:f pi}
\end{equation}
and the complex coefficients $\bar\alpha_{\vec{k}}$ are given by $\bar\alpha_{\vec{k}}=\bar{A}_{\vec{k}}+i\bar{B}_{\vec{k}}$, where the real and imaginary parts are Gaussian independent variables with zero mean and variance $\langle|\bar\alpha_{\vec{k}}|^{2}\rangle/2$. The two-point functions of the field and the field velocity are given by
\begin{equation}
\langle \tilde{\xi}(\vec{q})\,\tilde{\xi}^{*}(\vec{p})\rangle=\frac{a_{i}^{2}\langle|\alpha_{\vec{q}}|^{2}\rangle\delta^{(3)}(\vec{q}-\vec{p})}{\left(2\pi\right)^{3}\omega_{\vec{q}}}\,,
\label{eq:two-point xi}
\end{equation}
\begin{equation}
\langle \tilde{\pi}(\vec{q})\,\tilde{\pi}^{*}(\vec{p})\rangle=\frac{a_{i}^{4}\omega_{\vec{q}}}{\left(2\pi\right)^{3}}\,\langle|\alpha_{\vec{q}}|^{2}\rangle\delta^{(3)}(\vec{q}-\vec{p}),
\label{eq:two-point xi dot}
\end{equation}
where the dimensionless two-point function $\langle|\alpha_{\vec{q}}|^{2}\rangle$ will be chosen in a specific way depending on the power spectrum we want for the scalar field. 

In this work, we will consider two types of spectra so as to get a sense of the possibilities: thermal and hard cutoff. In the former case, $\langle|\alpha_{\vec{q}}|^{2}\rangle$ is given by the Bose-Einstein occupation number at temperature $T$,
\begin{equation}
\langle|\alpha_{\vec{q}}|^{2}\rangle=\frac{1}{e^{\omega_{\vec{q}}/T}-1}\,.
\label{eq:Bose-Einstein}
\end{equation}
On the other hand, for the hard cutoff spectrum, we will define $\langle|\alpha_{\vec{q}}|^{2}\rangle$ as
\begin{equation}
\langle|\alpha_{\vec{q}}|^{2}\rangle=
\begin{cases}
z^{2}/2 & \text{if\,\,\,\,} |\vec{q}|\leq Q\\\\
0 & \text{otherwise}
\end{cases}
\label{eq:cutoff spectrum}
\end{equation}
where $z$ is a constant that parametrizes the size of the fluctuations. We note that  $z=1$ would be the choice for standard vacuum fluctuations; but we allow for a possible enhancement above that with $z>1$, as we do not assume the field begins in its vacuum. Also, $Q$ is the cutoff wavenumber, which one anticipates to be associated with a microscopic scale in any fundamental theory.

Finally, the energy density of the scalar field at the initial time is given by
\begin{equation}
\rho(\eta_{i},\vec{r})=\frac{1}{2a_{i}^{4}}\xi'^{2}+\frac{1}{2a_{i}^{4}}(\vec{\nabla}\xi)^{2}+\frac{1}{2a_{i}^{2}}\tilde{M}^2\,\xi^{2}-{\langle H\rangle_i \over a_{i}^3}\xi\,\xi'+V_{\text{min}}\,,
\label{eq:rho}
\end{equation}
where $V_{\rm min}$ is the minimum of the potential\footnote{For inflation, one can take $V_{\rm min}\approx0$, while for slow contraction models (ekpyrosis) one often has a negative value for $V_{\rm min}$, as we will discuss.} and 
\begin{equation}
\tilde{M}^2=M^2+\langle H^2\rangle_i\,.
\end{equation}
Note the additional terms appearing in $\rho$ arise from the switch from $\delta\varphi$ to $\xi$ with $\delta\varphi=\xi/a$. In these terms, the Hubble rates appear in some kind of ensemble average. In the first appearance, it is $\langle H^2\rangle_i$, while in the second appearance, it is $\langle H\rangle_i$. Since there is no a priori reason in this framework for the universe to be initially expanding ($H_i>0$) or contracting ($H_i<0$), we take the initial ensemble average to be
\begin{equation}
\langle H\rangle_i=0\,.
\end{equation}
On the other hand, the ensemble average of the {\em square} of the Hubble parameter cannot be trivially dismissed and is written as
\begin{equation}
    \langle H^2\rangle_i={8\pi G\over 3}\langle\rho_{\rm tot}\rangle_i
\end{equation}
where $\langle\rho_{\rm tot}\rangle$ is the ensemble average of the total energy density. We do not include a curvature term $\langle k\rangle/a_i^2$ here because there is no a priori reason for the universe to be spherical or hyperbolic and so the average $\langle k\rangle$ can vanish (this also means we ignore the $\langle k\rangle$ term in (\ref{eq:effective mass})).
Typically, the correction from this $\langle H^2\rangle_i$ term in $\tilde{M}^2$ in the energy density $\rho$ is tiny and ignored for reasons as follows: in the case of a thermal spectrum, one normally estimates
\begin{equation}
\langle\rho_{\rm tot}\rangle={g_*\,\pi^2\over30}T^4\,,
\end{equation}
where $g_*$ is the effective number of degrees of freedom. Feeding this back into the above expression for $\rho$, one sees that the effective mass term is 
\begin{equation}
\tilde{M}^2\to M^2+{4\pi^3g_*\over45\,M_{p}^2}T^4\,.
\end{equation}
This new correction is negligible at moderate to low temperatures as it is Planck mass suppressed. However, at extremely high temperatures, namely $T\gg\sqrt{M\,M_{p}}$, it will be a significant correction to the mass.  It will turn out that these very high temperatures are not quite where the distribution is maximal, so this will not play a central role. So for the most part we will ignore such a correction, but we will indicate its effects in later sections as needed.

Let us also note that the ensemble-averaged Ricci scalar in (\ref{eq:effective mass}) is given by
\begin{equation}
\langle R\rangle=8\pi G\left(\langle\rho_{\rm tot}\rangle-3\langle p_{\rm tot}\rangle\right)\,,
\label{eq:Ricci ensemble average}
\end{equation}
where $p_{\rm tot}$ denotes the total pressure. In the case of a thermal spectrum, both $\langle\rho_{\rm tot}\rangle$ and $\langle p_{\rm tot}\rangle$ are exponentially suppressed in the limit of low temperatures ($T\ll M$), so (\ref{eq:Ricci ensemble average}) becomes negligible. 
For high temperatures ($T\gg M$) we have $\langle p_{\rm tot}\rangle\approx\langle\rho_{\rm tot}\rangle/3$, so there is a cancellation. The leading non-zero correction at high temperatures is 
\begin{equation}
\langle R\rangle\approx    \frac{2\pi}{3}{M^2T^2\over M_p^2}
\end{equation}
This is therefore negligible compared to the particle's mass $M^2$ in eq.~(\ref{eq:effective mass}), unless one is at Planck temperatures $T\sim M_p$.

So for any sub-Planckian temperatures (which is our focus), we will ignore this term in the effective mass in equation (\ref{eq:effective mass}). Moreover, as previously stated, we will take $\langle k\rangle=0$, so the angular frequencies defined in (\ref{eq:omega}) will be approximated as 
\begin{equation}
\omega_{\vec{k}}\approx\sqrt{M^{2}+\vec{k}^{2}}\,.
\label{eq:omega final}
\end{equation}



\section{Probability estimates}
\label{sec:probability estimates}

\subsection{Dominant Mode}

Our starting point is the fact that successful\footnote{By successful we mean able to smooth and flatten the universe to the desired degree.} inflation and ekpyrosis require a certain degree of homogeneity of the scalar field in some region of space. A simplified version of this condition is to consider that the energy density in this region is dominated by a single mode of the field. Let $\Lambda$ be the wavelength of this mode, and let $\hat{\rho}$ be the spatially averaged energy density over, say, 1/2 of the wavelength. If this region of space is approximately homogeneous and isotropic, the local Hubble rate $H_{\rm loc}$ should be given by the Friedmann equation,
\begin{equation}
H_{\rm loc}^{2}=\frac{8\pi G}{3}\hat\rho-\frac{k_{\rm loc}}{a_{i}^{2}}\,.
\label{eq:Friedmann}
\end{equation}
Since the amplitude and the phase of the dominant mode are stochastic, so are the energy density and the Hubble radius. We will say that inflation or slow contraction is successful if the field is sufficiently homogeneous on scales comparable to the horizon, or, in other words, if the Hubble radius is sufficiently small as compared to the scalar field wavelength:
\begin{equation}
|H_{\rm loc}|^{-1}<\Lambda/\chi\,,
\label{eq:success Hubble}
\end{equation}
where $\chi$ is a dimensionless  number that characterizes the precise criteria for inflation or slow contraction to commence. According to the examples presented in Table 1 of Ref.~\cite{Ijjas:2024oqn}, we have $\chi\sim10$ for inflation. 
This means the wavelength of the scalar field is about ten times longer than the Hubble radius, and it is suggested that slightly more adverse conditions (shorter field wavelengths) would lead to unsuccessful inflation.
For ekpyrosis, the table has $\chi\sim1$ (although to our knowledge, it could occur for smaller values too).

Furthermore, the inflationary initial conditions in that example are pretty close to flat FLRW, as indicated by the $\mathcal{O}(1)$ values of the (dimensionless) Weyl curvature and the Chern-Pontryagin invariant.  On the other hand, the initial conditions for the slow contraction example presented in the table have large values of the (dimensionless) Weyl curvature and the Chern-Pontryagin invariant. Even so, slow contraction successfully smooths and flattens the universe in that simulation, starting with a scalar field wavelength of the order of the horizon. 
In our work, we will not directly make use of the Weyl curvature or Chern-Pontryagin invariant. These quantities contain additional information, as the dimensionless versions involve ratios of the (square) of the Riemann curvature tensor to the (square) of the averaged Hubble parameter. 
In our simplifed analysis, these quantities are comparable, though in principle they do not have to be. Nevertheless, we use $\chi$ as a useful parameter to measure the difficulty in different phases commencing, and in particular we will take $\chi$ significantly larger for inflation.

\begin{figure}[t!]
\centering
\includegraphics[width=0.62\textwidth]{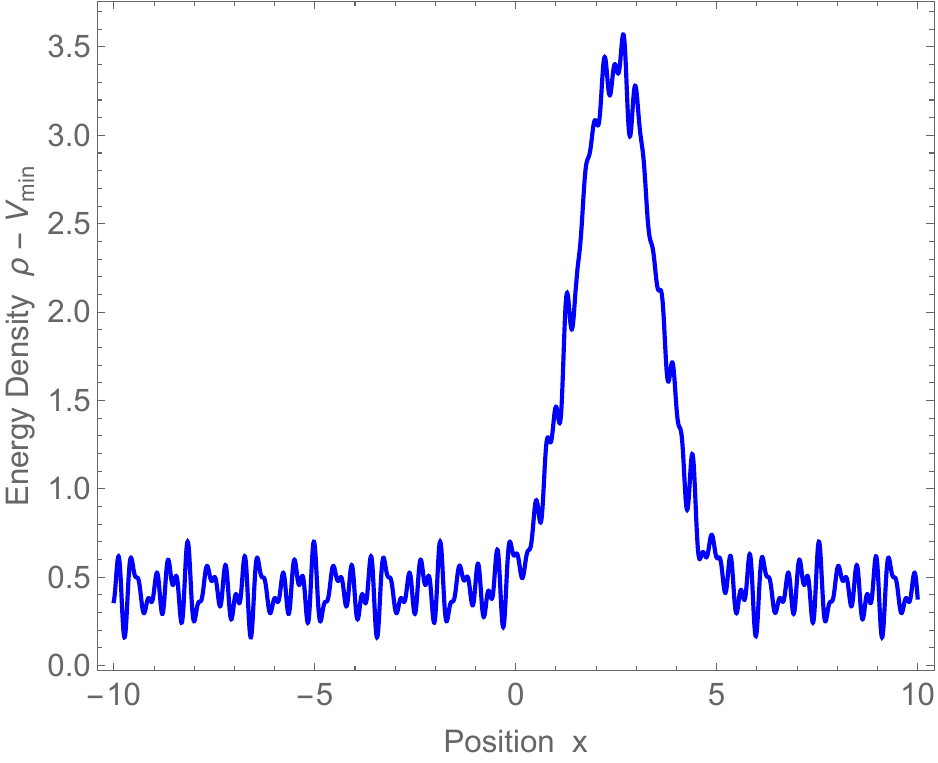}
\includegraphics[width=0.62\textwidth]{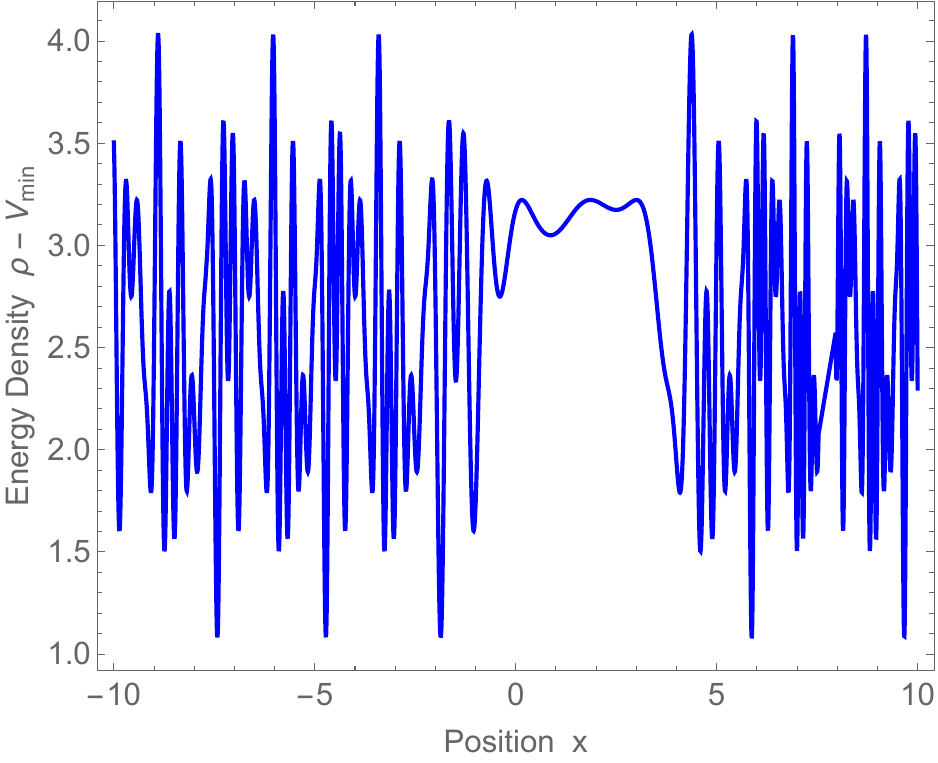}
\caption{Illustration of an interesting density field drawn from some distribution. 
The top figure is relevant for low temperatures/low cutoff: typically, the density is low. However, there is a chance of a rare fluctuation, as indicated here with the tall peak, that may satisfy the criteria for inflation or slow contraction (\ref{eq:success rho},\,\ref{eq:lamvalue},\,\ref{eq:condition UV modes}). The width of the peak is of order $\Lambda/4$. We note that for ekpyrosis the typical value of $\rho$ is close to $V_{\rm min}$, which is negative.
The bottom figure is relevant for high temperatures/high cutoff: typically, the density is large and dominated by UV fluctuations. However, there is a chance of a rare smooth region, as indicated here with the relatively low fluctuation region, that may also satisfy the criteria for inflation or slow contraction.}
\label{fig:figure random}
\end{figure}

The combination of (\ref{eq:Friedmann}) and (\ref{eq:success Hubble}) yields
\begin{equation}
\hat{\rho}>\frac{3}{8\pi G}\left(\frac{\chi^{2}}{\Lambda^{2}}+\frac{k_{\rm loc}}{a_{i}^{2}}\right)\equiv\rho_{*}\,.
\label{eq:success rho}
\end{equation}
Furthermore, we set $\Lambda/\chi$ to the be desired Hubble length for each theory, i.e., 
\begin{equation}
    \Lambda_{\rm inf}/\chi_{\rm inf}=H_{\rm inf}^{-1}\,\,\,\,\mbox{or}
    \,\,\,\,\,
    \Lambda_{\rm ekp}/\chi_{\rm ekp}=|H_{\rm ekp}|^{-1}
\label{eq:lamvalue}\end{equation} 
for inflation or slow contraction/ekpyrosis, respectively. With this in mind, the above criteria is that the stochastic local Hubble parameter obey $H_{\rm loc}\geq H_{\rm inf}$ or
$|H_{\rm loc}|\geq |H_{\rm ekp}|$. 
Typically, if this occurs, it will be in some isolated part of space surrounded by smaller fluctuations at low temperatures/low cutoff that do not satisfy this criteria; this is illustrated in top Figure \ref{fig:figure random}.
Or, if this occurs, it will be in some isolated part of space surrounded by larger fluctuations at high temperatures/high cutoff that do not satisfy this criteria; this is illustrated in bottom Figure \ref{fig:figure random}.
In practice, we will find that these rare events will be exponentially unlikely, and therefore, when this criteria is satisfied, it will take place with $H_{\rm loc}$  close to $H_{\rm inf}$ or $H_{\rm ekp}$. Our task in this section is to compute the probability that this will happen, that is, the probability associated to (\ref{eq:success rho}) (we will also need to address the issue of UV modes in the next subsection).

In the subsequent analysis, we will ignore the curvature term $k_{\rm loc}/a_{i}^{2}$ in (\ref{eq:success rho}).  We do not expect it to be relevant for our purposes, especially in the case of a closed universe, where this term has an upper limit. 

We note that in the case of inflation, this rare over-density will presumably not appear to actually inflate from the perspective of a distant observer. But nevertheless it can inflate from the point of view of an observer within the over-density. The outside observer will only see a horizon; so there is no direct contradiction.

Since the over-density for inflation or ekpyrosis is very large, one expects this to be approximately spherically symmetric. Hence any vector or tensor modes may be ignorable. Therefore, in order to compute the energy density of the dominant mode, we will consistently use the field and field velocity expansions presented in the previous section, which are valid for an approximately FLRW metric. Using (\ref{eq:xi expansion}), we can write our single-mode field as
\begin{equation}
\xi(\eta_{i},\vec{r})=a_{i}\sqrt{\frac{2}{L^{3}\omega_{\vec{K}}}}\left[A_{\vec{K}}\cos(\vec{K}\cdot\vec{r})-B_{\vec{K}}\sin(\vec{K}\cdot\vec{r})\right]\,,
\label{eq:single mode xi}
\end{equation}
where $|\vec{K}|=2\pi/\Lambda$ is the wave number of the dominant mode and the dimensionless $A_{\vec{K}}$ and $B_{\vec{K}}$ variables are related to the barred ones as follows:
\begin{equation}
\bar{A}_{\vec{K}}=a_{i}\left(\frac{L}{2\pi}\right)^{3/2}A_{\vec{K}}\,,
\label{eq:barred and unbarred}
\end{equation}
and likewise for $B_{\vec{K}}$. Here, $L=n\Lambda$ (with $n$ a positive integer or half-integer) is the size of the region where the mode dominates. Since $A_{\vec{K}}$ and $B_{\vec{K}}$ are Gaussian with zero mean and variance $\langle|\alpha_{\vec{K}}|^{2}\rangle/2$, one can rewrite them as
\begin{equation}
A_{\vec{K}}=\sqrt{\frac{\langle|\alpha_{\vec{K}}|^{2}\rangle}{2}}\,r_{\vec{K}}\cos(\gamma_{\vec{K}})\,,
\label{eq:A rewritten}
\end{equation}
\begin{equation}
B_{\vec{K}}=\sqrt{\frac{\langle|\alpha_{\vec{K}}|^{2}\rangle}{2}}\,r_{\vec{K}}\sin(\gamma_{\vec{K}})\,,
\label{eq:B rewritten}
\end{equation}
\\
where $\gamma_{\vec{K}}=2\pi f_{\vec{K}}$ and $r_{\vec{K}}=\sqrt{-2\ln(1-g_{\vec{K}})}$, with $f_{\vec{K}}$, $g_{\vec{K}}$ uniformly distributed between 0 and 1. With these redefinitions, (\ref{eq:single mode xi}) reads
\begin{equation}
\xi(\eta_{i},\vec{r})=a_{i}\sqrt{\frac{\langle|\alpha_{\vec{K}}|^{2}\rangle}{L^{3}\omega_{\vec{K}}}}\,r_{\vec{K}}\cos(\vec{K}\cdot\vec{r}+\gamma_{\vec{K}})\,.
\label{eq:single mode xi final}
\end{equation}
Similarly, the field velocity and the field gradient can be shown to be
\begin{equation}
\xi'(\eta_{i},\vec{r})=a_{i}^{2}\sqrt{\frac{\omega_{\vec{K}}\langle|\alpha_{\vec{K}}|^{2}\rangle}{L^{3}}}\,r_{\vec{K}}\sin(\vec{K}\cdot\vec{r}+\gamma_{\vec{K}})\,,
\label{eq:single mode pi}
\end{equation}
\begin{equation}
\vec{\nabla}\xi(\eta_{i},\vec{r})=a_{i}^{2}\vec{K}\,\sqrt{\frac{\langle|\alpha_{\vec{K}}|^{2}\rangle}{L^{3}\omega_{\vec{K}}}}\,r_{\vec{K}}\sin(\vec{K}\cdot\vec{r}+\gamma_{\vec{K}})\,.
\label{eq:single mode gradient}
\end{equation}
Finally, inserting these expressions in (\ref{eq:rho}), one finds
\begin{equation}
\rho(\eta_{i},\vec{r})=\frac{\omega_{\vec{K}}\left(A_{\vec{K}}^{2}+B_{\vec{K}}^{2}\right)}{L^{3}}-\frac{\vec{K}^{2}\left(A_{\vec{K}}^{2}+B_{\vec{K}}^{2}\right)}{\omega_{\vec{K}}L^{3}}\cos(2\vec{K}\cdot\vec{r}+2\gamma_{\vec{K}})+V_{\text{min}}\,.
\label{eq:single mode rho}
\end{equation}
where we have suppressed the $\langle H^2\rangle_i$ term for ease of presentation; but we shall return to its consequences in later sections.

Let us now parametrize the spatial average of $\rho$ as 
\begin{equation}
\hat{\rho}=\gamma\,\rho_{\text{max}}\,, 
\end{equation}
where
\begin{equation}
\rho_{\text{max}}=\frac{A_{\vec{K}}^{2}+B_{\vec{K}}^{2}}{L^{3}}\left(\omega_{\vec{K}}+\frac{\vec{K}^{2}}{\omega_{\vec{K}
}}\right)+V_{\text{min}}
\label{eq:rho max}
\end{equation}
is the local maximum value of $\rho$, as can be easily seen from (\ref{eq:single mode rho}),
and $\gamma$ is an $\mathcal{O}(1)$ number. Now, inequality (\ref{eq:success rho}) reads
\begin{equation}
A_{\vec{K}}^{2}+B_{\vec{K}}^{2}>\left(\frac{3\chi^{2}}{4\gamma}M_{p}^{2}\,\vec{K}^{2}-8\pi^{3}V_{\text{min}}\right)\frac{n^{3}\,\omega_{\vec{K}}}{|\vec{K}|^{3}\left(\omega_{\vec{K}}^{2}+\vec{K}^{2}\right)}\equiv R_{\vec{K}}^{2}\,.
\label{eq:success condition}
\end{equation}
Therefore, the problem reduces to the following question: what is the probability that $\alpha_{\vec{K}}=A_{\vec{K}}+iB_{\vec{K}}$ will fall outside a circle of radius $R_{\vec{k}}$ in the complex plane? This can be easily calculated as the joint probability density function of $A_{\vec{K}}$ and $B_{\vec{K}}$ is the product of two Gaussian distributions with zero mean and variance $\langle|\alpha_{\vec{K}}|^{2}\rangle/2$:\\
\begin{equation}
P\left(A_{\vec{K}}^{2}+B_{\vec{K}}^{2}>R_{\vec{K}}^{2}\right)=\frac{1}{\pi\langle|\alpha_{\vec{K}}|^{2}\rangle}\iint\limits_{A_{\vec{K}}^{2}+B_{\vec{K}}^{2}>R_{\vec{K}}} dA_{\vec{K}}\,dB_{\vec{K}}\,\exp\left(-\frac{A_{\vec{K}}^{2}+B_{\vec{K}}^{2}}{\langle|\alpha_{\vec{K}}|^{2}\rangle}\right)=\exp\left(-\frac{R_{\vec{K}}^{2}}{\langle|\alpha_{\vec{K}}|^{2}\rangle}\right)\,.
\label{eq:probability 1}
\end{equation}

\subsection{UV Modes}

We also have to demand the ultraviolet (UV) modes be energetically subdominant: 
\begin{equation}
\hat{\rho}_{_{UV}}<\rho_{*}\,. 
\label{eq:condition UV modes}
\end{equation}
We first estimate the spatially averaged energy density of these modes as
\begin{equation}
\hat{\rho}_{_{UV}}=\frac{1}{L^{3}}\sum_{|\vec{k}|>|\vec{K}|}\left(A_{\vec{k}}^{2}+B_{\vec{k}}^{2}\right)\,\omega_{\vec{k}}\,.
\label{eq:rho UV modes}
\end{equation}
Now, by the central limit theorem, $\hat{\rho}_{UV}$ should be approximately normally distributed, so the probability of (\ref{eq:condition UV modes}) is
\begin{equation}
P\left(\hat{\rho}_{_{UV}}<\rho_{*}\right)=\frac{1}{2}-\frac{1}{2}\erf\left(\frac{\langle\hat{\rho}_{_{UV}}\rangle-\rho_{*}}{\sqrt{2}\,\sigma}\right)\,,
\label{eq:probability 2}
\end{equation}
where $\langle\hat{\rho}_{_{UV}}\rangle$ and $\sigma^{2}$ are, respectively, the mean and the variance of $\hat{\rho}_{_{UV}}$. Using (\ref{eq:rho UV modes}), the former can be easily shown to be
\begin{equation}
\langle\hat{\rho}_{_{UV}}\rangle=\frac{1}{2\pi^{2}}\int_{0}^{\infty}dk\,k^{2}\sqrt{k^{2}+M^{2}}\,\langle|\alpha_{\vec{k}}|^{2}\rangle\,,
\label{eq:mean rho uv}
\end{equation}
while the latter is
\begin{equation}
\sigma^{2}=\langle\hat{\rho}_{_{UV}}^{2}\rangle-\langle\hat{\rho}_{_{UV}}\rangle^{2}=\frac{1}{2\pi^{2}L^{3}}\int_{0}^{\infty}dk\,k^{2}\left(k^{2}+M^{2}\right)\langle|\alpha_{\vec{k}}|^{2}\rangle^{2}\,.
\label{eq:variance rho uv}
\end{equation}

\subsection{Total probability}

Now we can finally calculate the total probability as the product of (\ref{eq:probability 1}) and (\ref{eq:probability 2}):
\begin{equation}
P=\frac{1}{2}\left[1-\erf\left(\frac{\langle\hat{\rho}_{_{UV}}\rangle-\rho_{*}}{\sqrt{2}\,\sigma}\right)\right]\exp\left(-\frac{R_{\vec{K}}^{2}}{\langle|\alpha_{\vec{K}}|^{2}\rangle}\right)\,.
\label{eq:total probability}
\end{equation}
This is our main result. It is an estimate of the probability that the initial state for inflation or ekpyrosis will be sufficiently homogeneous on scales comparable to the Hubble radius, assuming that the universe at that time is dominated by a free scalar field. Our result holds for an arbitrary power spectrum of the field, which is directly related to $\langle|\alpha_{\vec{k}}|^{2}\rangle$ via (\ref{eq:two-point xi}).

\subsection{Summary of Parameters}

As it stands, the total probability (\ref{eq:total probability}) depends on several parameters. Let us list them here: 
\begin{itemize}
\item $M$: mass of the scalar field.
\item $\Lambda$: wavelength of the dominant mode.
\item $V_{\text{min}}$: minimum of the potential energy density.
\item $\chi$: parametrizes the degree of homogeneity. Bigger $\chi$ means higher homogeneity (see equation (\ref{eq:success Hubble})).
\item $n$: number of wavelengths of the dominant mode in the nearly homogeneous region of space we are considering.
\item $\gamma$: indicates how much the spatially averaged energy density deviates from its mean value.
\item $T$: temperature in the case of thermal spectrum.
\item $Q$: UV cutoff in the case of hard-cutoff spectrum.
\item $z^2$: variance of the Gaussian variables $A_{\vec{k}}$ and $B_{\vec{k}}$ in the case of hard-cutoff spectrum.
\end{itemize}
The precise specification of $\chi$, $n$ and $\gamma$ will turn out to be irrelevant for our purposes, but the reader should keep in mind that they are order 1 or 10 (except for $\chi_{\rm ekp}$, which could be smaller). The key parameters will be the first three in the list above, and we will pick their values in agreement with the choices made in \cite{Kist:2022mew,Ijjas:2024oqn}, where the authors explore the smoothing capabilities of inflation and slow contraction using numerical relativity simulations. The order of magnitude of these parameters is shown in Planck units in Table \ref{table:parameters}. \\
\begin{table}[h!]
\centering
\begin{tabular}{|c ||c | c | c || }
\hline
 & $|H|/M_p$ & $M/M_p$ & $V_{\text{min}}/M_p^4$\\ [2ex] 
 \hline
Inflation & $10^{-5}\,\,$ & $10^{-4}\,\,$ & $0$\\ [2ex] 
 \hline
$\,$ Ekpyrosis$\,\,$ & $\,10^{-61}\,\,$ & $\,10^{-26}\,\,$ & $-10^{-55}$ \\[2ex]
 \hline
 \end{tabular}
\caption{Values of the key parameters Hubble rate $H$, field mass $M$, and potential minimum $V_{\rm min}$ (all in Planck units) that we will use to evaluate the probability in Eq.~(\ref{eq:total probability}). We are adopting most of these parameters from Refs.~\cite{Ijjas:2024oqn},\,\cite{Kist:2022mew}, with the addition of our choice for $H_{\rm inf}$. We note that this value of $M$ for inflation is perhaps a little higher than in some simple estimates, but it will suffice for our purposes. We do note that if the inflaton potential flattens at large field values, then this value for $M$ is in fact possible.}
\label{table:parameters}
\end{table}

For inflation, the potential near its minimum is
\begin{equation}
V_{\rm inf}(\varphi)\approx{1\over2}M_{\rm inf}^2(\delta\varphi)^2\,.
\end{equation}
At large field values for inflation, the shape can deviate away from this, but we shall use this quadratic approximation even at large values for simplicity.

For the case of ekpyrosis, note that the wavelength of the dominant mode is comparable to the Hubble radius today, $H_{0}^{-1}\sim10^{61}L_{p}$. Moreover, $M$ and $V_{\text{min}}$ in that case are computed from one of the potentials used in \cite{Kist:2022mew} (see also \cite{Tukhashvili:2023itb}):
\begin{equation}
V_{\text{ekp}}\left(\varphi\right)=-\frac{1}{2}V_{0}e^{-\varphi/m}\left[1+\tanh\left(\frac{\varphi-\varphi_{*}}{m_{*}}\right)\right]+V_{\text{DE}}\,,
\label{eq:ekpyrotic potential}
\end{equation}
with $V_{0}=\left(0.1/8\pi\right)M_{p}^{2}H_{0}^{2}$, $V_{\text{DE}}=V_{0}$, $m=\left(0.1/\sqrt{8\pi}\right)M_{p}$, $m_{*}=0.5\,m$ and $\varphi_{*}=-160\,m$. The minimum of the potential is located at
\begin{equation}
\varphi_{\text{min}}=\varphi_{*}+\frac{m_{*}}{2}\ln\left(\frac{2m}{m_{*}}-1\right)\approx-3.2M_{p}\,,
\label{eq:phi min}
\end{equation}
and so
\begin{equation}
V_{\text{min}}=V\left(\varphi_{\text{min}}\right)=V_{0}+V_{0}\left(\frac{2m}{m_{*}}-1\right)^{-\frac{m_{*}}{2m}}\left(\frac{m_{*}}{2m}-1\right)\,e^{-\frac{\varphi_{*}}{m}} \,.
\label{eq:Vmin}
\end{equation}
Hence we can estimate
\begin{equation}
V_{\rm min}\sim-10^{66}M_{p}^{2}H_{0}^{2}\sim -10^{-55}M_{p}^{4}
\end{equation}
and we note that this is {\em negative}. The presence of this minimum is necessary for slow contraction to end.

The mass of small fluctuations of the field about this minimum is
\begin{equation}
M_{\text{ekp}}=\sqrt{\frac{\left(\frac{2m}{m_{*}}-1\right)^{-\frac{m_{*}}{2m}}\left(m_{*}-2m\right)^{2}\exp\left(-\frac{\varphi_{*}}{m}\right)V_{0}}{2m_{*}m^{3}}}\sim10^{35}H_{0}\sim10^{-26}M_{p}\,.
\label{eq:mass ekpyrotic field}
\end{equation}
Expanding around the minimum, the ekypyrosis potential is then
\begin{equation}
V_{\rm ekp}(\varphi)\approx V_{\rm min}+{1\over2}M_{\rm ekp}^2\,(\delta\varphi)^2\,.
\end{equation}
Although the potential flattens at large field values, this leading order estimate will suffice.

In the following subsections, we will evaluate the total probability (\ref{eq:total probability}) for both inflation and ekpyrosis in the cases of a thermal and a hard-cutoff power spectrum of the scalar field. Before we do that, let us use (\ref{eq:success condition}) and the information in Table \ref{table:parameters} to simplify the exponent $R_{\vec{K}}^{2}/\langle|\alpha_{\vec{K}}|^{2}\rangle$ in (\ref{eq:total probability}).
For inflation, we have $V_{\text{min}}=0$, and so
\begin{equation}
\text{Inflation:}\,\,\,\,R_{\vec{K}}^{2}=\frac{3n^{3}\chi^{4}}{16\pi^2\gamma}\frac{\sqrt{1+\left(\frac{\chi}{2\pi}\right)^{2}\left(\frac{M}{H}\right)^{2}}}{2+\left(\frac{\chi}{2\pi}\right)^{2}\left(\frac{M}{H}\right)^{2}}\left(\frac{M_{p}}{H}\right)^{2}\,.
\label{eq:exponent for inflation}
\end{equation}
In the case of slow contraction, note that $|V_{\text{min}}|\gg M_{p}^{2}\vec{K}^{2}\sim (L_{p}/\Lambda)^{2}M_{p}^{4}\sim 10^{-122}M_{p}^{4}$, so we can safely ignore the first term on the right-hand side of (\ref{eq:success condition}). Note also that $M_{\text{ekp}}\gg |\vec{K}|\sim H_{0}\sim 10^{-61}M_{p}$, so $\omega_{\vec{K}}\approx M_{\text{ekp}}$. Taking this into account, we get
\begin{equation}
\text{Ekpyrosis:}\,\,\,\,R_{\vec{K}}^{2}\approx-\frac{n^{3}\chi^{3}V_{\text{min}}}{M H^{3}}\,.
\label{eq:exponent for ekpyrosis}
\end{equation}
For later convenience, we will split the total probability into its two contributions by taking the natural logarithm of (\ref{eq:total probability}):
\begin{equation}
-\ln\left(P_{s}\right)=\frac{R_{\vec{K}}^{2}}{\langle|\alpha_{\vec{K}}|^{2}\rangle}-\ln\left[\frac{1}{2}-\frac{1}{2}\erf\left(\frac{\langle\hat{\rho}_{_{UV}}\rangle-\rho_{*}}{\sqrt{2}\,\sigma}\right)\right]\,,
\label{eq:total probability log}
\end{equation}
where the subindex $s$ will denote either inflation (``$s=\text{inf}\,$'') or ekpyrosis (``$s=\text{ekp}$'') and the right-hand side is to be computed with the corresponding parameters in each case. The bigger the right-hand side, the lower the probability. Recall that the first term encodes the probability that the energy density of the dominant mode is greater than $\rho_{*}$, while the second term corresponds to the probability that the UV modes are subdominant.


\section{Results for a thermal spectrum}
\label{sec:results thermal spectrum}
Given the broad range of mass scales in the problem (the mass of the inflaton is 22 orders of magnitude above the mass of the ekpyrotic field), we will estimate the probability (\ref{eq:total probability log}) in three regimes:\\

(i). Low temperatures: $T\ll M_{\text{ekp}}\sim10^{-7}$ GeV.\\

(ii). Intermediate temperatures: $M_{\text{ekp}}\ll T\ll M_{\text{inf}}\sim 10^{15}$ GeV.\\

(iii). High temperatures: $M_{\text{inf}}\ll T\lesssim M_{p}\sim 10^{19}$ GeV.\\

In the limit $T\ll M$, one can show that expressions (\ref{eq:mean rho uv}) and (\ref{eq:variance rho uv}) for the mean and the variance of the energy density of the UV modes reduce to
\begin{equation}
\langle\hat{\rho}_{_{UV}}\rangle\approx\frac{M^{5/2}T^{3/2}}{\sqrt{8\pi^3}}\,e^{-M/T}\,,
\label{eq:mean rho uv low T}
\end{equation}

\begin{equation}
\sigma^{2}\approx\frac{M^{7/2}T^{3/2}H^{3}}{\sqrt{8\pi^3}\,n^{3}\chi^{3}}\,e^{-M/T}\,.
\label{eq:variance rho uv low T}
\end{equation}
\\
In the opposite limit, $T\gg M$, one finds
\begin{equation}
\langle\hat{\rho}_{_{UV}}\rangle\approx\frac{\pi^{2}}{30}T^{4}\,,
\label{eq:mean rho uv high T}
\end{equation}

\begin{equation}
\sigma^{2}\approx\frac{2\left[\pi^{4}-90\zeta(5)\right]T^{5}H^{3}}{15\pi^{2}n^{3}\chi^{3}}\approx0.055\frac{T^{5}H^{3}}{n^{3}\chi^{3}}\,,
\label{eq:variance rho uv high T}
\end{equation}
\\
where $\zeta$ is the Riemann zeta function. With the last four equations, as well as with (\ref{eq:exponent for inflation}) and (\ref{eq:exponent for ekpyrosis}), we are ready to evaluate (\ref{eq:total probability log}) in the three regimes listed above.

\subsection{Low temperatures}

The range of temperatures we will consider in this subsection is the following: $T\ll M_{\text{ekp}}\sim10^{-7}$ GeV.
In this limit, since $\langle\hat{\rho}_{_{UV}}\rangle$ is exponentially suppressed and $\rho_{*}$ is not, the argument of the error function in (\ref{eq:total probability log}) is dominated by the latter term, and the logarithm on the right-hand side becomes vanishingly small. This means that, at such low temperatures, the probability that the UV modes are energetically subdominant is almost 1. On the other hand, since $\langle|\alpha_{\vec{K}}|^{2}\rangle\approx\exp(-\omega_{\vec{K}}/T)$, we are led to

\begin{equation}
-\ln\left(P_{\text{inf}}\right)\approx C\,\left(\frac{M_{p}}{H_{\rm inf}}\right)^{2}\exp\left[\frac{\sqrt{M_{\text{inf}}^2+\left(\frac{2\pi H_{\rm inf}}{\chi_{\text{inf}}}\right)^2}}{T}\right]\,,
\label{eq:log prob inf low T}
\end{equation}

\begin{equation}
-\ln\left(P_{\text{ekp}}\right)\approx-\frac{n_{\rm ekp}^{3}\chi_{\rm ekp}^{3}V_{\text{min}}}{M_{\rm ekp} H_{\rm ekp}^{3}}\exp\left(\frac{M_{\text{ekp}}}{T}\right)\,,
\label{eq:log prob ekp low T}
\end{equation}
where 
\begin{equation}
C=\frac{3n_{\rm inf}^{3}\chi_{\rm inf}^{4}}{16\pi^2\gamma_{\rm inf}}\frac{\sqrt{1+\left(\frac{\chi_{\rm inf}}{2\pi}\right)^{2}\left(\frac{M_{\rm inf}}{H_{\rm inf}}\right)^{2}}}{2+\left(\frac{\chi_{\rm inf}}{2\pi}\right)^{2}\left(\frac{M_{\rm inf}}{H_{\rm inf}}\right)^{2}}\sim 1.
\label{eq:C}
\end{equation}
The fact that $M_{\text{inf}}, H_{\text{inf}}\gg M_{\text{ekp}}$ implies that the probability of the ekpyrotic setting in the low-temperature regime is overwhelmingly higher than that of the inflationary one.

\subsection{Intermediate temperatures}

The range of temperatures we will consider in this subsection is the following: $M_{\text{ekp}}\ll T\ll M_{\text{inf}}\sim 10^{15}$ GeV.
In this case, the probability for inflation will also be given by (\ref{eq:log prob inf low T}), but the ekpyrotic one will change. As before, let us start by evaluating the argument of the error function in (\ref{eq:total probability log}). For ekpyrosis, we find that $\langle\hat{\rho}_{_{UV}}\rangle\sim T^{4}\gg M_{\rm ekp}^{4}$ and $\rho_{*}\sim M_{p}^{2}H_{\rm ekp}^{2}\sim10^{-18}M_{\rm ekp}^{4}$. Therefore, the argument of the error function is dominated by the first term and the logarithm is huge. This is expected as the UV modes become more important at high temperatures. It is not hard to show that the second term on the right-hand side of (\ref{eq:total probability log}) can be approximated as $(n_{\text{ekp}}\Lambda_{\text{ekp}}T)^{3}$. On the other hand, the Bose-Einstein variance for ekpyrosis will approximately be $\langle|\alpha_{\vec{K}}|^{2}\rangle\approx T/M$. Taking all this into account, one finds

\begin{equation}
-\ln\left(P_{\text{inf}}\right)\approx C\,\left(\frac{M_{p}}{H_{\rm inf}}\right)^{2}\exp\left[\frac{\sqrt{M_{\text{inf}}^2+\left(\frac{2\pi H_{\rm inf}}{\chi_{\text{inf}}}\right)^2}}{T}\right]\,,
\label{eq:log prob inf intermediate T}
\end{equation}

\begin{equation}
-\ln\left(P_{\text{ekp}}\right)\approx \frac{n_{\text{ekp}}^{3}\chi_{\text{ekp}}^{3}}{H_{\rm exp}^{3}}\left(0.98\,T^{3}-d\frac{V_{\text{min}}}{T}\right)\,.
\label{eq:log prob ekp intermediate T}
\end{equation}
where
\begin{equation}
d=\left(1+\mathcal{O}\left(\langle H^2\rangle_i\over M_{\rm ekp}^2\right)\right)^{-1}\,.
\end{equation}
In this parameter $d$, we have indicated the order of the correction from the $\langle H^2\rangle_i$ term in eq.~(\ref{eq:rho}). 
As mentioned earlier, we have $\langle H^2\rangle_i\sim T^4/M_p^2$.; so we have $d\approx 1$ for all temperatures except the very high temperature regime $T\gtrsim\sqrt{M_{\rm ekp}\,M_p}$, where $d$ becomes small.

\begin{figure}[t!]
\centering
\includegraphics[width=0.85\textwidth]{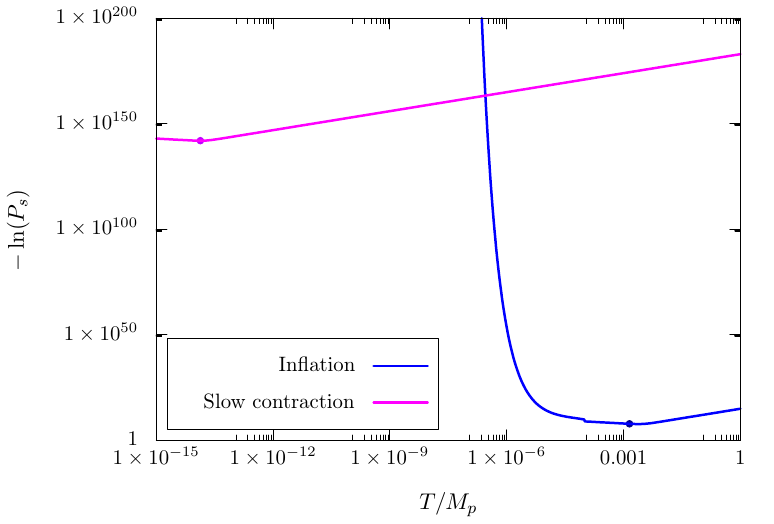}
\caption{Right-hand side of the inflationary probabilities (blue) at intermediate and high temperatures, (\ref{eq:log prob inf intermediate T}) and (\ref{eq:log prob inf high T}), and the probability for slow contraction (magenta) at high temperatures, (\ref{eq:log prob ekp high T}). The inflationary initial conditions become much more likely than the ekpyrotic ones when the temperature is high enough. For the plot, we have set all the parameters in the equations to 1. The two dots indicate the minimum of each function, corresponding to the maximum probability of inflation and slow contraction (see equations (\ref{eq:Tmax inf}) to (\ref{eq:log prob max ekp})).}
\label{fig:figure prob}
\end{figure}

Once again, in terms of probability, the ekpyrotic setting is greatly enhanced with respect to the inflationary one for most of the temperatures in this range. However, as shown in Figure \ref{fig:figure prob}, the situation is reversed for $T\gtrsim 10^{-3}M_{\text{inf}}\sim 10^{12}$ GeV.

\subsection{High temperatures}

The range of temperatures we will consider in this subsection is the following: $M_{\text{inf}}\ll T\lesssim M_{p}\sim 10^{19}$ GeV.
Using the same arguments as in the other two regimes, we get

\begin{equation}
-\ln\left(P_{\text{inf}}\right)\approx\frac{n_{\text{inf}}^{3}\chi_{\text{inf}}^{3}}{H_{\rm inf}^{3}}\left[0.98\,T^{3}+\frac{3 D}{8\pi}\,\frac{M_{p}^{2}H_{\rm inf}^{2}}{T}\right]\,,
\label{eq:log prob inf high T}
\end{equation}

\begin{equation}
-\ln\left(P_{\text{ekp}}\right)\approx \frac{n_{\text{ekp}}^{3}\chi_{\text{ekp}}^{3}}{H_{\rm exp}^{3}}\left(0.98\,T^{3}-d\frac{V_{\text{min}}}{T}\right)\,,
\label{eq:log prob ekp high T}
\end{equation}
where
\begin{equation}
D=\frac{1}{\gamma_{\rm inf}}\,\frac{1+\left(\frac{\chi_{\rm inf}}{2\pi}\right)^{2}\left(\frac{M_{\rm inf}}{H_{\rm inf}}\right)^{2}}{2+\left(\frac{\chi_{\rm inf}}{2\pi}\right)^{2}\left(\frac{M_{\rm inf}}{H_{\rm inf}}\right)^{2}}\,\tilde{d},\,\,\,\,\,\mbox{with}\,\,\,\,\tilde{d}=\left(1+\mathcal{O}\left(\langle H^2\rangle_i\over M_{\rm inf}^2\right)\right)^{-1}\,.
\end{equation}
Here $D\sim 1$ for all but the high temperature regime $T\gtrsim \sqrt{M_{\rm inf}/M_p}$, where we have to track the $\langle H^2\rangle_i/M_{\rm inf}^2\sim T^4/(M_p^2M_{\rm inf}^2)$ correction in $\tilde{d}$ coming from eq.~(\ref{eq:rho}), leading to $D$ becoming small.

In this case, since the mass of the inflaton is within a few orders of magnitude of the Planck mass, we are considering temperatures quite close to this scale. The second term inside the brackets on the right-hand side of (\ref{eq:log prob inf high T}) and (\ref{eq:log prob ekp high T}) is negligible, so the ratio of these two expressions is approximately given by $\ln(P_{\text{inf}})/\ln(P_{\text{ekp}})\sim(H_{\text{ekp}}/H_{\text{inf}})^{3}$, which is very small. Therefore, the inflationary setting is much more likely than the ekpyrotic one in the high-temperature regime.

\subsection{Maximum Probability}

From these last results, one can also find the temperature that maximizes the probability. For inflation, this temperature is

\begin{equation}
T_{\rm inf}^{\rm (max)}\approx\left(0.0406\,D\,M_{p}^{2}H_{\rm inf}^{2}\right)^{1/4}\,,
\label{eq:Tmax inf}
\end{equation}
\\corresponding to a probability

\begin{equation}
-\ln\left(P_{\text{inf}}^{\rm (max)}\right)\approx0.355\,n_{\rm inf}^{3}\chi_{\rm inf}^{3} D^{3/4}\left(\frac{M_{p}}{H_{\rm inf}}\right)^{3/2}\,.
\label{eq:log prob max inf}
\end{equation}
\\
For slow contraction, we have

\begin{equation}
T_{\rm ekp}^{\rm (max)}\approx\left(-\frac{V_{\rm min}}{2.94}\right)^{1/4}\,,
\label{eq:Tmax ekp}
\end{equation}
\\corresponding to a probability

\begin{equation}
-\ln\left(P_{\text{ekp}}^{\rm (max)}\right)\approx1.75\,n_{\rm ekp}^{3}\chi_{\rm ekp}^{3}\frac{\left(-V_{\rm min}\right)^{3/4}}{H_{\rm ekp}^{3}}\,.
\label{eq:log prob max ekp}
\end{equation}
\\
Neglecting terms of order one in the temperatures, we find $T_{\text{inf}}^{(\text{max})}\sim10^{-3}M_{p}\sim 10^{16}$ GeV and $T_{\text{ekp}}^{(\text{max})}\sim 10^{-14}M_{p}\sim 10^{5}$ GeV. Regarding the probabilities, we note that the last factor in (\ref{eq:log prob max inf}) is $(M_{p}/H_{\rm inf})^{3/2}\sim 10^{7}$, while in (\ref{eq:log prob max ekp}) we have $(-V_{\rm min})^{3/4}/H_{\rm ekp}^{3}\sim 10^{141}$. So the maximum probabilities can be roughly estimated as
\begin{equation}
-\ln\left(P_{\text{inf}}^{\rm (max)}\right)\sim 10^7,\,\,\,\,\,\,-\ln\left(P_{\text{ekp}}^{\rm (max)}\right)\sim 10^{141}\,.
\end{equation}
Therefore, even the maximum probabilities are very small. But it is interesting to note that the maximum probability is much, much smaller for the case of ekpyrosis compared to inflation.


\subsection{Other Degrees of Freedom}

In addition to the single scalar field (the inflaton or ekpyrotic field), there could be other degrees of freedom that can contribute to the UV, or thermal, modes. These could be the degrees of freedom of the Standard Model (or if we track both fields, it could be the other field).
In this case we should update the above formulas in (\ref{eq:log prob ekp intermediate T}), (\ref{eq:log prob inf high T}), and (\ref{eq:log prob ekp high T}) by the replacement
\begin{equation}
    0.98T^3\to 0.98\, \hat{g}_*\,T^3\,,
\end{equation}
where $\hat{g}_*$ is a kind of effective number of degrees of freedom, whereby $g_*=1$ (for each boson) that has mass $M_j\ll T$. This does not alter our final qualitative results for the maximum probability, but it just introduces some factor of $g_*$, which we anticipate to be at most $g_*\sim 100$ or so.

\section{Results for a hard-cutoff spectrum}
\label{sec:results cutoff spectrum}

For the hard-cutoff spectrum (\ref{eq:cutoff spectrum}), the mean and the variance of the energy density of the UV modes (expressions (\ref{eq:mean rho uv}) and (\ref{eq:variance rho uv}), respectively) reduce to

\begin{equation}
\langle\hat{\rho}_{_{UV}}\rangle=\frac{z^2}{32\pi^2}\left[Q\left(M^{2}+2Q^{2}\right)\sqrt{M^{2}+Q^{2}}-M^{4}\arcsinh\left(\frac{Q}{M}\right)\right]\,,
\label{eq:mean rho uv hard cutoff}
\end{equation}

\begin{equation}
\sigma^{2}=\frac{Q^{3}z^{4}H^{3}}{8\pi^{2}n^{3}\chi^{3}}\left(\frac{M^{2}}{3}+\frac{Q^2}{5}\right)\,.
\label{eq:variance rho uv hard cutoff}
\end{equation}

In order for the first term on the right-hand side of (\ref{eq:total probability log}) to be finite, we will consider $Q>(2\pi/\chi_{\rm inf}) H_{\text{inf}}$. Let us start now with the slow contraction case, where we have $Q\gg M_{\rm ekp}$. Consequently, the condition $\langle\hat{\rho}_{_{UV}}\rangle\gg\rho_{*}$ will be met if

\begin{equation}
z^{2}\gg\frac{6\pi M_{p}^{2}H_{\rm ekp}^{2}}{Q^{4}}\,.
\label{eq:conditions hard cutoff 1}
\end{equation}
Since $Q>(2\pi/\chi_{\rm inf}) H_{\text{inf}}$, the condition (\ref{eq:conditions hard cutoff 1}) is guaranteed if

\begin{equation}
z^{2}\gg z_{*}^{2}\equiv\frac{3\chi_{\rm inf}^{4} M_{p}^{2}H_{\rm ekp}^{2}}{8\pi^{3}H_{\rm inf}^{4}}\,.
\label{eq:conditions hard cutoff 2}
\end{equation}

For the parameters we are using, we have $z_{*}^{2}\sim 10^{-100}$. Therefore, the condition (\ref{eq:conditions hard cutoff 1}) is always satisfied and the dominant term in the argument of the error function is $\langle\hat{\rho}_{_{UV}}\rangle$. This allows us to use the same arguments as in the high-temperature limit above to get

\begin{equation}
-\ln\left(P_{\text{ekp}}\right)\approx\frac{n_{\text{ekp}}^{3}\chi_{\text{ekp}}^{3}}{H_{\rm ekp}^{3}}\left(\frac{5Q^{3}}{64\pi^{2}}-\frac{2V_{\text{min}}}{z^{2}M_{\text{ekp}}}\right)\,.
\label{eq:log prob ekp cutoff}
\end{equation}

Let us turn now to the probability of inflation. The quantity $-\ln(P_{\rm inf})$ is maximized in the regime where $\langle\hat{\rho}_{_{UV}}\rangle$ dominates over $\rho_{*}$, and this happens for $Q\gg M_{\rm inf}$. In this limit, the result is

\begin{equation}
-\ln\left(P_{\text{inf}}\right)\approx\frac{n_{\text{inf}}^{3}\chi_{\text{inf}}^{3}}{H_{\rm inf}^{3}}\left[\frac{5Q^{3}}{64\pi^{2}}+\frac{2C}{n_{\rm inf}^{3}\chi_{\rm inf}^{2}}\,\frac{M_{p}^{2}H_{\rm inf}}{z^{2}}\right]\,.
\label{eq:log prob inf cutoff}
\end{equation}

Since this is always smaller than (\ref{eq:log prob ekp cutoff}), the inflationary conditions are more likely than the ekpyrotic ones for any value of $Q$.


\section{Summary and conclusions}
\label{sec:conclusions}
One of the main problems attributed to the inflationary paradigm is the apparent failure of the inflaton to smooth and flatten the universe starting from the highly inhomogeneous and anisotropic conditions expected to emerge from a big bang. Recent studies supported by numerical relativity simulations claim that slow contraction, unlike inflation, is a robust smoothing mechanism that can account for the homogeneity, isotropy and flatness of the universe observed today on large scales, even if the initial state is far from FLRW \cite{Cook:2020oaj, Ijjas:2023dnb, Garfinkle:2023vzf,Ijjas:2024oqn}.

The performance of inflation and slow contraction is tested separately in two different types of simulations (``inflation simulations'' and ``slow contraction simulations'') in \cite{Ijjas:2024oqn}. Although this is a comparative study, the scales involved in these two types of simulations are vastly different at the initial time: the analogue of the Hubble radius and the wavelength of the scalar field modes in the slow contraction simulations are more than 60 orders of magnitude greater than in the inflation ones. Given these disparate initial conditions, we wonder how likely it is for the inflaton and ekpyrotic fields to be relatively homogeneous on those scales, as it might be the case that slow contraction is the preferred smoothing mechanism at the cost of requiring a much more unlikely initial configuration.

In this paper, we have addressed this question under the assumption that the universe is dominated by a free scalar field at the beginning of inflation or slow contraction. Our main result, equation (\ref{eq:total probability}), is an estimate of the probability that this field is homogeneous on scales comparable to the Hubble radius at that time. It depends crucially on the mass and the power spectrum of the field, as well as on the desired scale of homogeneity. We have evaluated the total probability using typical values of the parameters employed in \cite{Ijjas:2024oqn} for a thermal and a hard-cutoff spectrum. Roughly speaking, the inflationary initial conditions are more likely than the ekpyrotic ones when the cutoff energy scale (the temperature in the case of a thermal spectrum) is above the inflationary Hubble rate.

As a final comment, let us mention that, in order to fully assess the likelihood of the initial conditions, one would have to incorporate the probabilties associated with the geometric configuration of spacetime at the initial time. According to \cite{Ijjas:2024oqn} and the previous studies in this series of papers, successful inflation requires very small values of the Weyl curvature and the Chern-Pontryagin invariant relative to ekpyrosis. We leave the question of how much this disfavors the inflationary scenario in terms of the probability of the initial conditions for future work. We also postpone the analysis of more detailed contributions that come from the imposition of the slow-roll conditions in the case of inflation or fast-roll for slow contraction.


\section{Acknowledgments}

M.~P.~H. is supported in part by National Science
Foundation grants PHY-2310572 and PHY-2419848.
D.~J.-A. is supported in part by National Science 
Foundation grants PHY-2110466 and PHY-2419848.
We thank Alan Guth for helpful discussion.

\appendix

\section{Geometry of the initial-time hypersurface}
\label{sec:appendix}
Even if the main focus of the paper is to study the probability associated to the homogeneity of the initial conditions for the scalar field, in this appendix we provide a characterization of the geometry of the initial-time hypersurface in terms of its mean curvature (the analogue of the Hubble rate in a FLRW universe) and the power spectrum of the metric perturbations. This simple analysis may pave the way towards a probabilistic description of the geometry of the initial state, which should be ultimately used to complement the estimates derived in the main text for the energy content. 

For simplicity, we will consider the regime of linear gravity in a Minkowski background. Even if this is not the actual situation that one would expect near the Big Bang, the analysis presented here will allow us to see that the typical wavelength of the scalar field modes is extremely short compared to the mean curvature scale. Therefore, homogeneity of the field on larger scales is indeed a very unlikely condition.

The scalar field fluctuations distort the Minkowski metric as $g_{\mu\nu}=\eta_{\mu\nu}+h_{\mu\nu}$. In this case, since the background Einstein tensor is zero, the perturbed Einstein equations read
\begin{equation}
\delta G_{\mu\nu}=8\pi G\left(\bar{T}_{\mu\nu}+\delta T_{\mu\nu}\right)\,,
\label{eq:EE}
\end{equation}
where $\bar{T}_{\mu\nu}$ is the background energy-momentum tensor. At the linear level, the 00 component of the perturbed Einstein tensor is given by
\begin{equation}
\delta G_{00}=\frac{1}{2}\left[-\nabla^{2}\left(h_{11}+h_{22}+h_{33}\right)+\partial_{i}\partial_{j}h_{ij}\right]\,.
\label{eq:G00}
\end{equation}
We will only focus on the scalar sector of the metric perturbations:\\
\begin{equation}
h_{\mu\nu}=
\begin{pmatrix}
2\phi & \partial_{1}B & \partial_{2}B & \partial_{3}B \\
\partial_{1}B & 2\psi+\partial_{1}^{2}E & 2\partial_{1}\partial_{2}E & 2\partial_{1}\partial_{3}E \\
\partial_{2}B & 2\partial_{1}\partial_{2}E & 2\psi+2\partial_{2}^{2}E & 2\partial_{2}\partial_{3}E \\
\partial_{3}B & 2\partial_{1}\partial_{3}E & 2\partial_{2}\partial_{3}E & 2\psi+2\partial_{3}^{2}E \\
\end{pmatrix}\,.
\end{equation}
\\
In the Newtonian gauge, the coordinates are chosen in such a way that $B$ and $E$ vanish. Therefore, the 00 component of (\ref{eq:EE}) in the Newtonian gauge is
\begin{equation}
\nabla^{2}\psi=4\pi G\rho\,.
\label{eq:equation for psi}
\end{equation}
\subsection{Power spectrum of the metric perturbations}
Let $\tilde{\psi}(\vec{k})$ and $\tilde{\rho}(\vec{k})$ be the Fourier transforms of $\psi$ and $\rho$, respectively:
\begin{equation}
\psi(\vec{r})=\int d^{3}k\,\tilde{\psi}(\vec{k})e^{i\vec{k}\cdot\vec{r}}\,,
\label{eq:ft psi}
\end{equation}
\begin{equation}
\rho(\vec{r})=\int d^{3}k\,\tilde{\rho}(\vec{k})e^{i\vec{k}\cdot\vec{r}}\,.
\label{eq:ft rho}
\end{equation}
Then, it follows from (\ref{eq:equation for psi}) that
\begin{equation}
\tilde{\psi}(\vec{k})=-4\pi G\,\frac{\tilde{\rho}(\vec{k})}{|\vec{k}|^{2}}\,.
\label{eq:h psi}
\end{equation}
Therefore, in order to calculate the power spectrum of $\psi$, $\langle|\tilde{\psi}(\vec{k})|^{2}\rangle$, we need the power spectrum of $\rho$. On can compute $\tilde{\rho}(\vec{k})$ as the sum of the Fourier transforms on the right-hand side of (\ref{eq:rho}), which can be found using the convolution theorem. For instance, the Fourier transform of $\xi^{2}$ is
\begin{equation}
\tilde{\xi^{2}}(\vec{k})=\int d^{3}q\,\,\tilde{\xi}(\vec{q})\,\tilde{\xi}(\vec{k}-\vec{q})\,.
\label{eq:ft xi squared}
\end{equation}
With this in mind, and setting $V_{\rm min}=0$, one finds
\begin{equation}
\tilde{\rho}(\vec{k})=\frac{1}{2}\int d^{3}q\left[\left(M^{2}+\vec{q}\cdot\left(\vec{q}-\vec{k}\right)\right)\tilde{\xi}(\vec{q})\tilde{\xi}(\vec{k}-\vec{q})+\tilde{\pi}(\vec{q})\tilde{\pi}(\vec{k}-\vec{q})\right]
\label{eq:ft rho 2}
\end{equation}
and the two-point function
\begin{equation}
\langle|\tilde{\rho}(\vec{k})|^{2}\rangle=\frac{\delta^{(3)}(\vec{0})}{4\left(2\pi\right)^{6}}\int d^{3}q\,\langle|\alpha_{\vec{q}}|^{2}\rangle\langle|\alpha_{\vec{k}-\vec{q}}|^{2}\rangle\left[2\omega_{\vec{q}}\omega_{\vec{k}-\vec{q}}+\left(M^{2}+\vec{q}\cdot\left(\vec{q}-\vec{k}\right)\right)\frac{2\omega_{\vec{q}}-\vec{k}^{2}}{\omega_{\vec{q}}\omega_{\vec{k}-\vec{q}}}\right]
\label{eq:ps rho}
\end{equation}
for $\vec{k}\neq\vec{0}$. In order to get this result, we have used Wick's theorem:
\begin{equation}
\begin{split}
\langle \tilde{\xi}(\vec{k}_{1})\tilde{\xi}(\vec{k}_{2})\tilde{\xi}(\vec{k}_{3})\tilde{\xi}(\vec{k}_{4})\rangle&=\langle \tilde{\xi}(\vec{k}_{1})\tilde{\xi}(\vec{k}_{2})\rangle\langle \tilde{\xi}(\vec{k}_{3})\tilde{\xi}(\vec{k}_{4})\rangle+\langle \tilde{\xi}(\vec{k}_{1})\tilde{\xi}(\vec{k}_{3})\rangle\langle \tilde{\xi}(\vec{k}_{2})\tilde{\xi}(\vec{k}_{4})\rangle+\\
&+\langle \tilde{\xi}(\vec{k}_{1})\tilde{\xi}(\vec{k}_{4})\rangle\langle \tilde{\xi}(\vec{k}_{2})\tilde{\xi}(\vec{k}_{3})\rangle\,,
\end{split}
\label{eq:Wick theorem}
\end{equation}
and likewise for $\tilde{\pi}(\vec{k})$. Equation (\ref{eq:ps rho}) can be finally used to compute the power spectrum of $\psi$ as 
\begin{equation}
\langle|\tilde{\psi}(\vec{k})|^{2}\rangle=\frac{16\pi^{2}\langle|\tilde{\rho}(\vec{k})|^{2}\rangle}{M_{p}^{4}\lvert\vec{k}\rvert^{4}}\,.
\label{eq:ps psi}
\end{equation}

\subsection{Mean curvature scale}

In addition to the power spectrum of the metric fluctuations, we will also be interested in the induced mean curvature scale $\Theta^{-1}$, which is analogous to the Hubble rate in a FLRW spacetime. This scale is given by 
\begin{equation}
\Theta^{-1}=\frac{\mathcal{K}}{3}\,,
\label{eq:H1}
\end{equation}
where $\mathcal{K}$ is the trace of the extrinsic curvature of the initial-time hypersurface. In order to find it, we first calculate the vector $n^{\mu}$ which is normal to this hypersurface: $n^{\mu}=\lambda\nabla^{\mu}t=\lambda g^{\mu 0}$. Imposing the normalization condition $n^{\mu}n_{\mu}=-1$, one finds that the proportionality factor is $\lambda=\pm\sqrt{1-2\phi}$, and so
\begin{equation}
n^{\mu}=\left(\mp\frac{1}{\sqrt{1-2\phi}},0,0,0\right)\,.
\label{eq:normal vector}
\end{equation}
The trace of the extrinsic curvature is then computed as $\mathcal{K}=\nabla_{\mu}n^{\mu}=\partial_{0}n^{0}+n^{0}\Gamma^{\mu}_{0\mu}$. Since
\begin{equation}
\Gamma^{0}_{00}=-\frac{\dot{\phi}}{1-2\phi}\,\,\,\,\,,\,\,\,\,\,\Gamma^{1}_{01}=\Gamma^{2}_{02}=\Gamma^{3}_{03}=\frac{\dot{\psi}}{1+2\psi}\,,
\label{eq:Christoffel symbols}
\end{equation}
one finally gets
\begin{equation}
\Theta^{-1}=\mp\frac{\dot{\psi}}{\left(1+2\psi\right)\sqrt{1-2\phi}}\,.
\label{eq:H2}
\end{equation}
It follows from (\ref{eq:equation for psi}) that the Fourier transform of $\dot{\psi}$ is
\begin{equation}
\tilde{\dot{\psi}}(\vec{k})=-4\pi\,\frac{\tilde{\dot{\rho}}(\vec{k})}{|\vec{k}|^{2}}\,.
\label{eq:f psi dot}
\end{equation}
Taking the time derivative of the Minkowski version of (\ref{eq:rho}) and using the equation of motion of the field, (\ref{eq:eom}), one finds that
\begin{equation}
\dot{\rho}=\dot{\xi}\nabla^{2}\xi+\xi_{,i}\dot{\xi}_{,i}\,,
\label{eq:rho dot}
\end{equation}
and from the convolution theorem we get 
\begin{equation}
\tilde{\dot{\rho}}(\vec{k})=-\int d^{3}q\,(\vec{k}\cdot\vec{q})\tilde{\xi}(\vec{q})\tilde{\pi}(\vec{k}-\vec{q})\,.
\label{eq:f rho dot}
\end{equation}
Using this expression, it is easy to show that the average of $\dot{\psi}$ (and consequently, that of $\Theta^{-1}$) is zero. Therefore, we will estimate the average scale of interest as 
\begin{equation}
H\equiv\sqrt{\langle\dot{\psi}^{2}\rangle}\,.
\label{eq:H}
\end{equation}
One can use (\ref{eq:f psi dot}) and (\ref{eq:f rho dot}) to show that
\begin{equation}
\langle\dot{\psi}^{2}\rangle=\frac{1}{4\pi^{4}M_{p}^{4}}\int d^{3}q\int d^{3}p\,\frac{\omega_{\vec{q}-\vec{p}}\,(\vec{q}\cdot\vec{p})^{2}}{\omega_{\vec{p}}\,|\vec{q}|^{4}}\,\langle|\alpha_{\vec{p}}|^{2}\rangle\langle|\alpha_{\vec{p}-\vec{q}}|^{2}\rangle\,.
\label{eq:average psi dot squared}
\end{equation}

\subsection{Numerical results for a thermal spectrum}

Here we will present the results for the power spectrum of the metric perturbations and the mean curvature scale, given by (\ref{eq:ps psi}) and (\ref{eq:H}), respectively. We focus on a thermal spectrum here for concreteness. To this end, the integrals in (\ref{eq:ps rho}) and (\ref{eq:average psi dot squared}) were solved numerically in a cubic lattice with side length $L$ and lattice spacing $\Delta x$. 
In the following examples, we chose $L=10M^{-1}$ and $\Delta x=0.1M^{-1}$ for a total of $N^{3}=10^{6}$ points in our cubic lattice.
Let us also clarify that the discrete version of the Dirac delta function appearing in (\ref{eq:ps rho}) is $\delta^{(3)}(\vec{0})\rightarrow L^{3}/(2\pi)^{3}$, and $\int d^{3}q\rightarrow\sum_{l}\sum_{m}\sum_{n}\left(2\pi/L\right)^{3}$.

\begin{figure}[h!]
\centering
\includegraphics[width=0.79\textwidth]{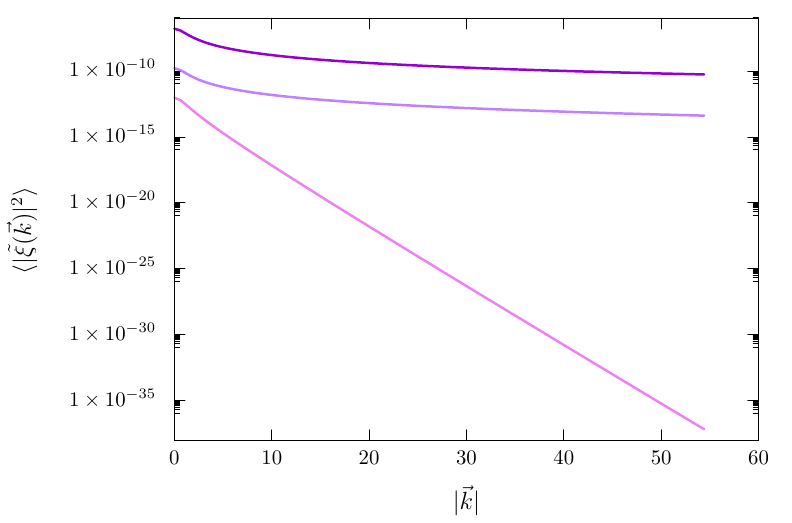}
\caption{Power spectrum of $\xi$ for the temperatures indicated in Table \ref{table:curvature scale}. Lighter shades of purple correspond to lower temperatures. In order to restore dimensions, the numbers on the horizontal axis must be multiplied by $M$, and the ones on the vertical axis, by $M_{p}^{2}/M^{6}$.}
\label{fig:figure xi 1}
\end{figure}

\begin{figure}[h!]
\centering
\includegraphics[width=0.79\textwidth]{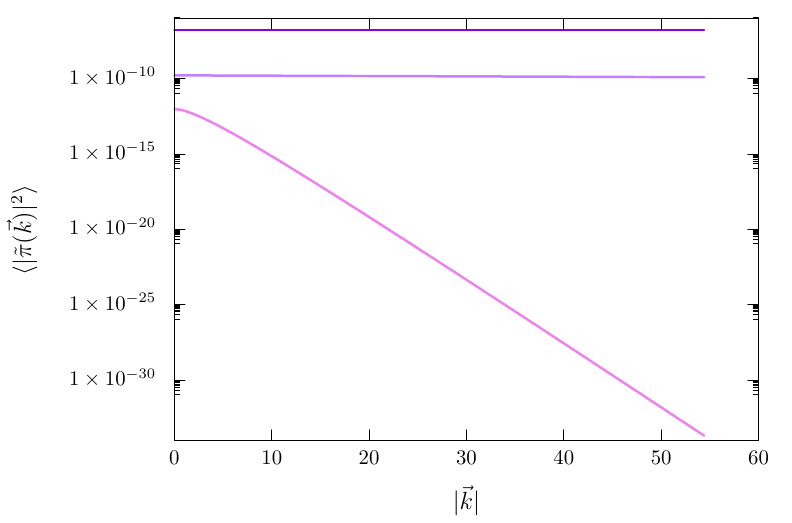}
\caption{Power spectrum of $\dot{\xi}$ for the temperatures indicated in Table \ref{table:curvature scale}. Lighter shades of purple correspond to lower temperatures. In order to restore dimensions, the numbers on the horizontal axis must be multiplied by $M$, and the ones on the vertical axis, by $M_{p}^{2}/M^{4}$.}
\label{fig:figure pi 1}
\end{figure}

\begin{figure}[h!]
\centering
\includegraphics[width=0.79\textwidth]{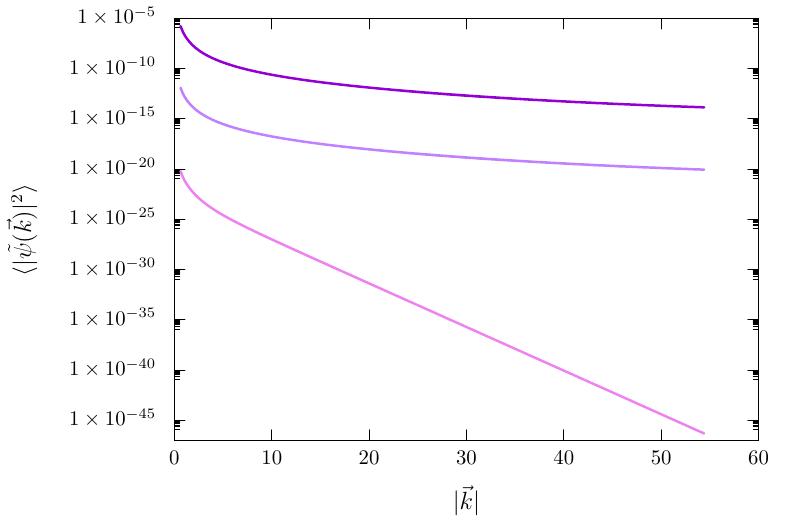}
\caption{Power spectrum of $\psi$ for the temperatures indicated in Table \ref{table:curvature scale}. Lighter shades of purple correspond to lower temperatures. In order to restore dimensions, the numbers on the horizontal axis must be multiplied by $M$, and the ones on the vertical axis, by $M^{-6}$. For the high-temperature cases, the spectrum decays very approximately as $|\vec{k}|^{-4}$.}
\label{fig:figure psi 1}
\end{figure}

If the scalar field is in a thermal state at temperature $T$, the two-point function of the $\alpha_{\vec{k}}$ coefficients is given by (\ref{eq:Bose-Einstein}),
which can be expressed as
\begin{equation}
\langle|\alpha_{\vec{k}}|^{2}\rangle=\frac{M}{2M_{p}}\left[\coth\left(\frac{\omega_{\vec{k}}}{2T}\right)-1\right]\,.
\label{eq:alpha thermal state}
\end{equation}
In the following, we show the results for a scalar field mass of $M=10^{-5}M_{p}$ and three different temperatures. 
In Table \ref{table:curvature scale}, we present the results for the induced curvature scale, as defined in (\ref{eq:H}), and in Figures \ref{fig:figure xi 1}, \ref{fig:figure pi 1} and \ref{fig:figure psi 1}, we plot, respectively, the power spectrum of $\xi$, $\dot{\xi}$ and $\psi$ in dimensionless units.\\
\begin{table}[h!]
\centering
\begin{tabular}{| c || c | c | c |}
\hline
$T/M_{p}$ & $1$ & $10^{-3}$ & $10^{-5}$ \\[2ex] 
 \hline
 $H/M_{p}$ & $1.28\times 10^{-7}$ & $1.086\times 10^{-10}$ & $2.60\times 10^{-15}$ \\ [2ex] 
 \hline
$H/T\sim\lambda/H^{-1}$ & $1.28\times 10^{-7}$ & $1.086\times 10^{-7}$ & $2.60\times 10^{-10}$\\[2ex]
 \hline
 \end{tabular}
\caption{Mean curvature scale for three different temperatures. In the third row, we show the ratio between the typical wavelength of the thermal field modes and the induced curvature scale, which should be the analog of the Hubble radius in a Friedmann-Robertson-Walker spacetime. If one extrapolates these results to the actual initial conditions near the Big Bang, the conclusion is that the scalar field is, typically, highly inhomogeneous on scales shorter than the horizon.}
\label{table:curvature scale}
\end{table}

\clearpage





\bibliography{inflationVSekpyrosis.bib}

\end{document}